\documentclass[conference]{IEEEtran}
\IEEEoverridecommandlockouts

\usepackage{pgfplots}
\pgfplotsset{compat=1.14}

\usepackage{adjustbox}

\usepackage{standalone}
\usepackage{tabularx,booktabs}
\newcolumntype{C}{>{\centering\arraybackslash}X} 
\usepackage{lipsum}
\usepackage{xcolor}
\usepackage{multirow}
\usepackage{tikz}
\usetikzlibrary{shapes,arrows}

\usepackage{amsmath,amsfonts,amsthm,bm} 

\usepackage{cite}
\usepackage{amsmath,amssymb,amsfonts}
\usepackage{algorithmic}
\usepackage{graphicx}
\usepackage{textcomp}
\usepackage{xcolor}
\def\BibTeX{{\rm B\kern-.05em{\sc i\kern-.025em b}\kern-.08em
    T\kern-.1667em\lower.7ex\hbox{E}\kern-.125emX}}

\definecolor{color1}{RGB}{0,119,187}
\definecolor{color4}{RGB}{51,187,238}
\definecolor{color3}{RGB}{0,153,136}
\definecolor{color6}{RGB}{255,112,67}
\definecolor{color2}{RGB}{204,51,17}
\definecolor{color5}{RGB}{238,51,119}

\usepackage{tikz}
\usetikzlibrary{shapes,arrows}

\begin{document}

\title{Study of Short-Term Personalized Glucose Predictive Models on Type-1 Diabetic Children}


\author{\IEEEauthorblockN{Maxime De Bois}
\IEEEauthorblockA{Universit\'{e} Paris Saclay\\
CNRS-LIMSI\\
Orsay, France\\
maxime.debois@limsi.fr}
\and
\IEEEauthorblockN{Moun\^{i}m A. El Yacoubi}
\IEEEauthorblockA{
T\'{e}l\'{e}com SudParis, Universit\'{e} Paris Saclay\\
SAMOVAR, CNRS\\
\'{E}vry, France\\
mounim.el\textunderscore yacoubi@telecom-sudparis.eu}
\and
\IEEEauthorblockN{Mehdi Ammi}
\IEEEauthorblockA{
Universit\'{e} Paris 8\\
Dept. of Computer Science\\
Saint-Denis, France\\
ammi@ai.univ-paris8.fr}}


\maketitle


\begin{abstract}
Research in diabetes, especially when it comes to building data-driven models to forecast future glucose values, is hindered by the sensitive nature of the data. Because researchers do not share the same data between studies, progress is hard to assess. This paper aims at comparing the most promising algorithms in the field, namely Feedforward Neural Networks (FFNN), Long Short-Term Memory (LSTM) Recurrent Neural Networks, Extreme Learning Machines (ELM), Support Vector Regression (SVR) and Gaussian Processes (GP). They are personalized and trained on a population of 10 virtual children from the Type 1 Diabetes Metabolic Simulator software to predict future glucose values at a prediction horizon of 30 minutes. The performances of the models are evaluated using the Root Mean Squared Error (RMSE) and the Continuous Glucose-Error Grid Analysis (CG-EGA). While most of the models end up having low RMSE, the GP model with a Dot-Product kernel (GP-DP), a novel usage in the context of glucose prediction, has the lowest. Despite having good RMSE values, we show that the models do not necessarily exhibit a good clinical acceptability, measured by the CG-EGA. Only the LSTM, SVR and GP-DP models have overall acceptable results, each of them performing best in one of the glycemia regions.
\end{abstract}
\begin{IEEEkeywords}
Glucose prediction, Feedforward Neural Network, Recurrent Neural Network, Long Short-Term Memory, Extreme Learning Machine, Support Vector Regression, Gaussian Process
\end{IEEEkeywords}

\section{Introduction}

With 1.5 milion inputed deaths in 2012 \cite{world2016global}, diabetes is one of the most threatening diseases of the modern world. Because of the non secretion of insulin (type-1 diabetes) or body resistance to its action (type-2 diabetes), diabetic people have trouble managing their blood glucose. When his blood glucose falls too low, the diabetic is said to be in a state of hypoglycemia, while, in the other hand, when it gets too high, we talk about hyperglycemia. Because both hypoglycemia and hyperglycemia have respectively short-term (e.g., exhaustion, coma, death) and long-term (e.g., cardiovascular diseases, blindness) complications, diabetic people need to maintain their blood glucose within an acceptable range.

Big advances have been made in the recent years to help diabetic people in their daily life. Continuous glucose monitoring (CGM) devices, such as the FreeStyle Libre \cite{olafsdottir2017clinical}, make possible to track the glucose level throughout days and nights without having to rely on the inconvenient use of lancets. Besides, in combination with CGM devices, we are witnessing a rise of coaching applications for diabetes such as the application  mySugr \cite{rose2013evaluating}, which has been approved by the Food \& Drug Administration (FDA) in the United States. In the other hand, bariatric surgery has been shown to induce a 10-year remission rate of type 2 diabetes of 36\% \cite{doi:10.1001/jamasurg.2014.2440}. Finally, since 2016, the first artificial pancreas, the MiniMed 670G, has been available in the United States \cite{minimed}. 


One of the biggest research area of interest is the prediction of future glucose values. For a diabetic patient, knowing accurately his future glycemia is undoubtedly valuable as it can be used to avoid getting into the hypo-/hyperglycemia ranges by modifying his behavior (e.g., by taking insulin shots or by eating sugar). 


Currently, the focus of glucose predictive models is heavily in favor of data-driven techniques, where the patient's glucose, carbohydrate (CHO) intake and insulin injection past values are used to forecast future glucose values. While the autoregressive (AR) model and its different variations are the most traditionally used models in the field \cite{sparacino2007glucose}, they have fallen out of favor for more complex regression models.

In particular, Zecchin \textit{et al.} showed that using meal information improves the forecasting of glucose by using a Feedforward Neural Network (FFNN) \cite{zecchin2012neural}. Recurrent Neural Networks (RNN) is a class of artificial neural network (ANN) made for time-series forecasting and have been used by Daskalaki \textit{et al.} to built a hypoglycemia early-warning system \cite{daskalaki2013early}. While RNN present some limitations (e.g., the vanishing gradient problem), novel types of the RNN cell have been engineered and recently tried out to predict future glucose values, such as the Long Short-Term Memory (LSTM) unit by Mirshekarian \textit{et al.} \cite{mirshekarian2017using}. Extreme Learning Machine (ELM) is another type of ANN that is quite popular nowadays thanks to its ability to provide relatively good generalization with close to no training time \cite{assadi2017estimation, jankovic2016deep}.

Meanwhile, models that use the kernel method, also known as the kernel trick, to map the initial space of observations into a higher dimension space, have shown interesting results when used to predict future glucose values. Georga \textit{et al.} and Khan \textit{et al.} investigated the usability of Support Vector Regression (SVR) in forecasting blood glucose \cite{georga2013multivariate,khan2017methods}. Besides, De Paula \textit{et al.} used Reinforcement Learning alongside with Gaussian Processes (GP) to predict future glucose levels and include the predictions in a decision-based system aiming at regulating blood glucose \cite{de2015controlling}.

However, those advances are hindered by several factors. Because of their sensitive nature, diabetes-related data used in studies are not made available to other researchers. This leads every research team to collect its own data and building their studies around them. Since most of the studies do not share the same data, they cannot compare to each other. The field needs comparative studies that give objective insights on the performances of the models. In 2012, Daskalaki \textit{et al.} led a study aiming at comparing two AR models and a FFNN, which ends up outperforming the former \cite{daskalaki2012real}. Meanwhile, in 2015, Zarkogianni \textit{et al.}, compared four different models they have investigated in previous studies, namely a FFNN model, a linear regression model, a Self-Organizing Map and a neuro-fuzzy network with wavelets \cite{zarkogianni2015comparative}. Finally, also in 2015, Georga \textit{et al.} evaluated hybrid glucose predictive models that combine regression models (SVR or GP) and feature ranking algorithms \cite{georga2015evaluation}.


Nowadays, it is still unclear how the most trending models relate to each other in terms of performances. In this study, we compare six of the most promising glucose predictive models, namely a FFNN, a RNN with LSTM units, an ELM neural network, a SVR model and two GP models. To address this goal, we first describe the data flow, from its simulation using the Type 1 Diabetes Metabolic Simulator software to the implementation of the models. Then, we discuss the results of the models obtained by evaluating them with two different metrics. Finally, we conclude by providing our takeaways and some guidelines for future studies.

\section{Methods}

\begin{figure}\scalebox{0.5}{
\tikzstyle{block} = [rectangle, draw, fill=blue!20, 
    text width=7em, text centered, rounded corners, minimum height=2em]
\tikzstyle{line} = [draw, -latex']

\begin{tikzpicture}[->]
    \node [fill=white] (parameters) at (0,3) {parameters};
    \node [block] (simulation) at (0,2) {Simulation};
    \node [block] (preprocessing) at (0,0) {Preprocessing};
    \node [block] (training) at (0,-2) {Training};
    \node [block] (evaluation) at (0,-4) {Evaluation};
    \node [fill=white] (rmse) at (-1,-5) {RMSE};
    \node [fill=white] (cg-ega) at (1,-5) {CG-EGA};
    
    \node[fill=white] (testing) at (3,-1) {testing set};
    \node[fill=white] (eval) at (-3,-1) {evaluation set};
    
    \path[line] (parameters) -- (simulation);
    \path[line] (simulation) -- (preprocessing) node[midway,fill=white]{time-series};
    \path[line] (preprocessing) -- (training) node[midway,fill=white]{training set};
    \path[-] (preprocessing) edge (3,0);
    \path[-] (3,0) edge (testing);
    \path[line] (testing) |- (3,-4) -- (evaluation);
    \path[-] (preprocessing) edge (-3,0);
    \path[-] (-3,0) edge (eval);
    \path[line] (eval) |- (-3,-2) -- (training);
    \path[line] (training) -- (evaluation) node[midway,fill=white]{model};
    \path[line] (evaluation) -- (rmse);
    \path[line] (evaluation) -- (cg-ega);
\end{tikzpicture}
}
\caption{Data flow diagram, from its simulation to the evaluation of the models.}
\label{fig:flowchart}
\end{figure}
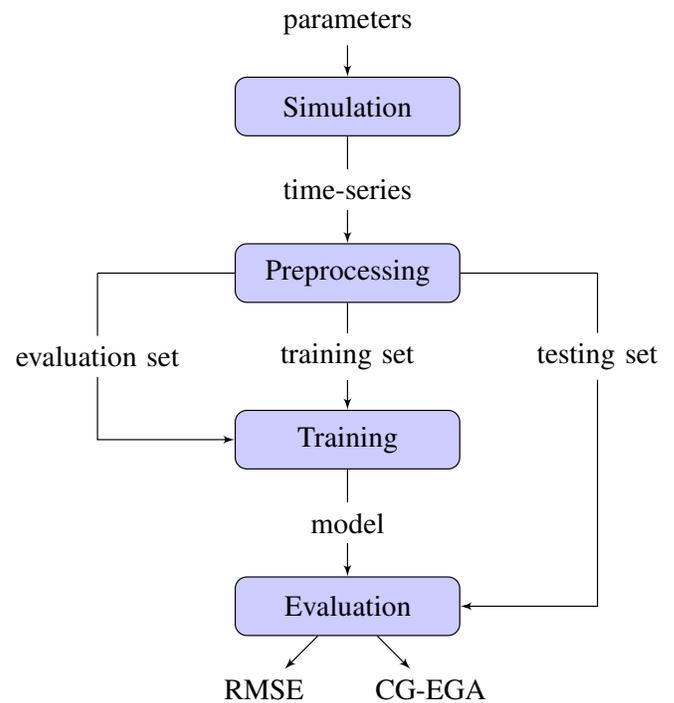

This section presents the whole methodology that has been used to compare the predictive models. First, we explain how we obtain our experimental data. Then, we go through the preprocessing of the data and the building of the predictive models. Finally, we discuss the evaluation metrics that have been used in this study. Figure \ref{fig:flowchart} illustrates this methodology.

\subsection{Data Simulation}

The T1DMS software \cite{t1dms} is a type-1 virtual diabetic patients simulator that has been accepted as a substitute to clinical testing by the FDA \cite{man2014uva} and that has been extensively used in the glucose prediction research field \cite{de2015controlling, zhao2018multiple, contreras2017personalized, zarkogianni2011insulin, zecchin2012neural}.


In this study, 10 \textit{in-silico} type-1 diabetic children (representing the most challenging diabetic population) underwent the following open-loop experiment:

\begin{itemize}
    \item 3-meal daily scenario, where the quantity and the timing of each meal have been randomized to account for the variability of real-life situations. More specifically, the timings have been sampled from normal distributions centered at 7h, 13h and 20h respectively and with a variance of 0.5. As for the quantities (CHO intakes in $g$), the normal distributions were centered at 40g, 85g and 60g with a variance of 0.5 times the mean quantity.
    \item An insulin bolus is taken at the start of a meal. The value of the bolus is taken uniformly between 0.7 and 1.3 times the patient's optimal bolus given his carbohydrate-to-insulin ratio.
    \item Basal insulin is constant and optimal (computed by the simulator).
\end{itemize}

Similar scenarios have been used in the past few years \cite{daskalaki2012real, zarkogianni2011insulin, zecchin2012neural}. The major difference between our simulation and others is the length of the simulation. While most simulations last only a few days, we simulated 29 days of data. This serves the purpose of enhancing the generalization of the models by avoiding overfitting.

In the end, for every patient, the simulation outputs three different time-series with a sampling frequency of 60 Hz: glucose values, carbohydrate intakes and insulin boluses over time.

\subsection{Data Preprocessing}

\subsubsection{Data Rearrangement}

\begin{figure*}
\tikzstyle{block} = [rectangle, draw, fill=blue!20, 
    text width=7em, text centered, rounded corners, minimum height=2em]
\tikzstyle{line} = [draw, -latex']

\begin{tikzpicture}[->]
    
    \node [fill=white] (t1) at (17,0) {\parbox{3em}{insulin}};
    \node [fill=white] (g1) at (17,0.5) {\parbox{3em}{CHO}};
    \node [fill=white] (CHO1) at (17,1.0) {\parbox{3em}{glucose}};
    \node [fill=white] (ins1) at (17,1.5) {\parbox{3em}{time}};
    \node [fill=white] (...1) at (10,0) {...};
    \node [fill=white] (...2) at (10,0.5) {...};
    \node [fill=white] (...3) at (10,1.0) {...};
    \node [fill=white] (...4) at (10,1.5) {...};

    \node [fill=white] (day1) at (2,1.85) {day 1};
    \node [fill=white] (day2) at (6,1.85) {day 2};
    \node [fill=white] (day28) at (14,1.85) {day 29};
    
    \node [red,fill=white] (day2) at (5.5,-1.1) {subset 1};
    \node [red,fill=white] (day28) at (13.5,-1.1) {subset 28};
    
    \path[-] (0,0) edge (9.5,0);
    \path[-] (0,0.5) edge (9.5,0.5);
    \path[-] (0,1.0) edge (9.5,1.0);
    \path[-] (0,1.5) edge (9.5,1.5);
    \path[line] (10.5,0) -- (t1);
    \path[line] (10.5,0.5) -- (g1);
    \path[line] (10.5,1.0) -- (CHO1);
    \path[line] (10.5,1.5) -- (ins1);
    
    \path[-] (0,-0.25) edge (0,1.75);
    \path[-] (4,-0.25) edge (4,1.75);
    \path[-] (8,-0.25) edge (8,1.75);
    \path[-] (12,-0.25) edge (12,1.75);
    \path[-] (16,-0.25) edge (16,1.75);
    
    \draw[dashed,red,very thick] (3,-0.75) rectangle (8,2.25);
    \draw[dashed,red,very thick] (11,-0.75) rectangle (16,2.25);
    
    
\end{tikzpicture}
\caption{Data rearrangement: from the 29-day long time-series, 28 subsets are created.}
\label{fig:rearrangement}
\end{figure*}
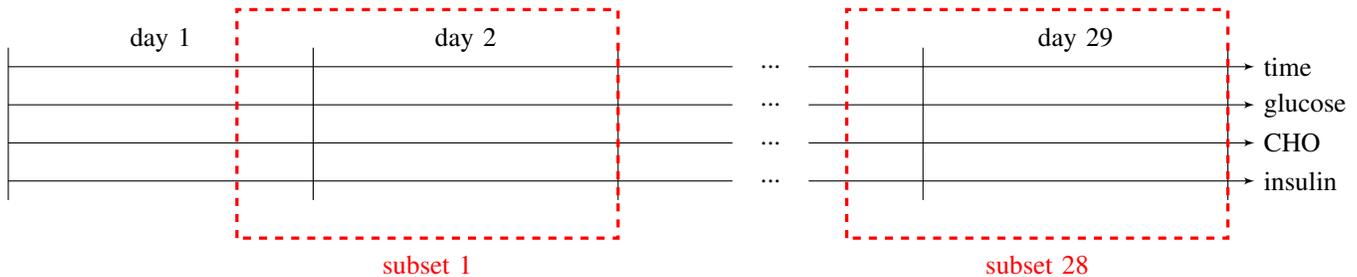

First, we divided the 29-day long time-series into day-long subsets. Then, we expanded every daily subset with the data of the previous day to account for the prediction horizon (30 minutes) and the data history (60 minutes) used in the models. Finally we discarded the first day as it is mostly used to warm up the simulator. We end up having 28 subsets of 1530 samples-long time-series. Figure \ref{fig:rearrangement} illustrates this process.

~

\subsubsection{3-way Data Splitting}\label{subsubsec:3way}

Half of the subsets, the training set, is used to train the models. A quarter, the evaluation set, is used to tune the hyperparameters of the models. Finally, the remaining quarter, the testing set, is used to evaluate the final model. We note that, even-though this kind of splitting is a common procedure in Machine Learning, it has not been used in the past in glucose prediction (a simpler training and testing sets split is relatively common, though). 

~

\subsubsection{Standardization}

The training set is standardized (zero-mean and unit variance) and the same transformation is then applied to the evaluation and testing sets.

\subsection{Glucose Predictive Models}
\label{sec:methods-glucose}

This subsection goes through all the models that have been implemented. The training and optimization of the models have been personalized to the patient.


~

\subsubsection{Feedforward Neural Network (FFNN)}

The network is made of 4 hidden layers of respectively 128, 64, 32 and 16 neurons. Apart from the last neuron, whose activation function is linear, every neuron uses the Scaled Exponential Linear Unit (SELU), which is the Exponential Linear Unit (ELU) with an optimized value of $\alpha$ \cite{klambauer2017self}. All the weights are initialized using a coarse model trained on one of the children. This serves the purpose of speeding up the training process of the remaining children. The model is fine-tuned to the given patient using the Adam optimizer with the Mean Squared Error loss function, mini-batches of size 1500, an initial learning rate of $10^{-3}$ and a decay of $10^{-4}$. To avoid the slight overfitting encountered during the training of the models, we add up a L2 penalty to the weights and made use of early stopping. Alpha-dropout (specific to the use of the SELU activation function) has been tried out with no perceptible improvement. Also, the SELU activation function, not used so far in glucose prediction studies, has shown to work really well with better training time and performances compared to more classical activation functions (e.g., tanh, ReLU).

~

\subsubsection{Long Short-Term Memory Recurrent Neural Network (LSTM)}

The LSTM network is made of one recurrent layer of 64 LSTM units. The network has been unrolled 60 times to account for an history of 60 minutes. As for the FFNN model, the weights are initialized to coarse values fitted on one of the children. The model is then fine-tuned using the Adam optimizer with the Mean Squared Error loss function, mini-batches of size 500 and an initial learning rate of $10^{-2}$. As for the amount of regularization used during the training, we added a L2 penalty to the weights ($10^{-4}$), used early stopping and recurrent dropout (rate of 0.5). Since an increase in the number of hidden layers or hidden neurons did not yield better results, we stuck to a rather simple network. 

~

\subsubsection{Extreme Learning Machine (ELM)}

ELM networks are quite simple to tune as we only need to adjust the number of neurons in the single hidden layer and their activation function \cite{huang2006extreme}. The logistic activation function seemed to be the one to work the best for us. We applied a L2 penalty (100.0) to the weights to reduce the impact of overfitting. While continuously adding more neurons improved the performances, we chose to stop at 20160 neurons (which is the number of training samples) as the increase in performance was not significant afterwards.

~

\subsubsection{Support Vector Regression (SVR)}

The SVR model has been implemented using the radial basis function (RBF) kernel (defined in Equation \ref{eqn:rbf}, with $x$ and $x'$ being two input vectors). The kernel's coefficient ($\gamma$) has been personalized and optimized by grid search from the initial $[10^{-2}, 10^1]$ range. The parameter $\epsilon$ models the $\epsilon$-tube within which no penalty is associated to the training loss function. While the penalty is also personalized and optimized within a specific range ($[10^{-5}, 10^{-2}]$), $\epsilon$ has been fixed to $10^{-3}$. Lower values of $\epsilon$ made the model unable to fit the training data while greater values yielded worse results.

\begin{equation}
\label{eqn:rbf}
k(x, x') = \exp(- \gamma \| x-x'\| ^2)
\end{equation}

~

\subsubsection{Gaussian Process (GP) Regression}

As for the SVR model, GP models are traditionally used with a RBF kernel in glucose prediction studies \cite{georga2015evaluation}. However, we found out that using a dot-product (DP) kernel (defined in Equation \ref{eqn:dotproduct}, with $x$ and $x'$ being two input vectors) was way more effective. Therefore, we implement both versions of Gaussian Process regression: one with a RBF kernel (GP-RBF) and one with a DP kernel (GP-DP).

\begin{equation}
\label{eqn:dotproduct}
k(x, x') = \sigma_0 ^ 2 + x \cdot x' 
\end{equation}

~

The RBF kernel has been fixed with a value of $\gamma$ of 0.5 as changing the values did not impact the results. As for the DP kernel, the value of $\sigma_0$ has also been fixed to a value of $0.01$ for the same reason.

In order to help our models to fit the training data, we added noise to the observations, represented by the value $\alpha$ that is added to the diagonal of the kernel matrix during fitting. More noise implies an easier fit but also worse results, so we personalized and optimized its value in the $[10^{-2}, 10^1]$ range.

\subsection{Performance Metrics}

We evaluate the performances of the models using nested cross-validation \cite{krstajic2014cross}, doing permutations of the training, evaluation and testing sets splitting (which is described in Section \ref{subsubsec:3way}). A first cross-validation between the training and the evaluation sets is used to tune the hyperparameters of the models. Then, the tuned and fitted models are evaluated with a second round of cross-validation with the testing set.

A lot of different evaluation metrics have been used throughout the years \cite{oviedo2017review}. We have settled down with the two most significant ones : the Root Mean Squared Error (RMSE) and the Continuous Glucose Error Grid Analysis (CG-EGA).

~

\subsubsection{Root Mean Squared Error}

The RMSE (defined in Equation \ref{eqn:rmse}, with $y$ being the true values and $\hat{y}$ being the predicted values) is the standard metric to measure the performance of a glucose predictive model.  It has the advantage of being a single value metric making comparison between models straightforward. It can also be used as the loss function during the training stage (e.g., when training neural networks). And, compared to other similar metrics (e.g., the Mean Absolute Error), it penalizes bigger errors more, which is preferable in glucose prediction since big errors, even when they are rare, can have disastrous consequences.

\begin{equation}
\label{eqn:rmse}    
RMSE(y,\hat{y}) = \sqrt{\frac{1}{n} \sum_{i=1}^n{(y_i - \hat{y}_i)^2}}
\end{equation}

~

\subsubsection{Continuous Glucose-Error Grid Analysis}

The CG-EGA, initially introduced to measure the clinical acceptability of Continuous Glucose Monitoring (CGM) devices \cite{kovatchev2004evaluating}, sees a lot of use in the evaluation of glucose predictive models \cite{oviedo2017review}. It is made of two different evaluation grids: the Point-Error Grid Analysis (P-EGA) and the Rate-Error Grid Analysis (R-EGA). With the P-EGA, depending on the true glucose value, predictions are assigned to clinical acceptability areas, from A to E (i.e. good to bad). As for the R-EGA, the idea is the same but we focus on rates of change instead of focusing on the point-values themselves. The CG-EGA is simply the Cartesian product of the P-EGA and the R-EGA. In order to appreciate the CG-EGA, it is simplified by giving each cell a measure of its clinical acceptability, depending of the true state of the patient's glycemia. A cell can then either contain accurate predictions (AP), benign errors (BE) or erroneous predictions (EP).

\section{Results}

The experimental results have been reported in Table \ref{table:results} which depicts, in terms of average RMSE and CG-EGA across the children, the performance of the six models described in section \ref{sec:methods-glucose}.

With the biggest RMSE and, in most glycemia areas in the CG-EGA, the biggest amount of EP, the ELM model comes out to be the worst in our study.

In terms of RMSE, the SVR, the FFNN and the GP-DP models stand out from the other models by making predictions the closest to the actual glucose values, with the GP-DP model being the best of the three.

As for the clinical acceptability of the models, the conclusions depend on the glycemia range. In the range of euglycemia (where the true glucose value is between 70 mg/dL and 180 mg/dL), all the models manage to make acceptable predictions with a minimum of 0.09\% (SVR and LSTM) and a maximum of 1\% (ELM) of EP. However, in the hypoglycemia range (true value below 70 mg/dL), the ELM and the GP-RBF models show clinically unacceptable results with a significant number of EP. In the hyperglycemia range, the ELM and the FFNN models show also unacceptable results with a high number of EP and BE. Overall, the LSTM, SVR and GP-DP models show stable good results across the whole CG-EGA table.

\begin{figure*}
\begin{adjustbox}{width=\textwidth}
\begin{tikzpicture}
\begin{axis}[
    xlabel={Time [minutes]},
    xlabel style={font=\tiny},
    ylabel={Glucose concentration [mg/dL]},
    ylabel style={font=\tiny},
    xmin=0, xmax=1440,
    ymin=25, ymax=325,
    legend pos=north west,
    legend style={font=\tiny},
    ytick={50,100,150,200,250,300},
    xticklabel style={font=\tiny},
    yticklabel style={font=\tiny},
    domain=0:1440,
]
\addplot[
    color=color1,
    line width=1.0pt,
    ]
    table {data/graph/elm_child-002_day_5.dat};
 \addplot[
    color=color2,
    line width=1.0pt,
    ]
    table {data/graph/gp_rbf_child-002_day_5.dat};
\addplot[
    color=color3,
    line width=1.0pt,
    ]
    table {data/graph/lstm_child-002_day_5.dat};
\addplot[
    color=color4,
    line width=1.0pt,
    ]
    table {data/graph/svm_child-002_day_5.dat};
\addplot[
    color=color5,
    line width=1.0pt,
    ]
    table {data/graph/ffnn_child-002_day_5.dat};
\addplot[
    color=color6,
    line width=1.0pt,
    ]
    table {data/graph/gp_dp_child-002_day_5.dat};
\addplot[
    color=black,
    line width=0.5pt,
    ]
    table {data/graph/base_child-002_day_5.dat};
\addplot[mark=none, black, samples=2, style=dashed,] {70.0};
\addplot[mark=none, black, samples=2, style=dashed,] {180.0};
\legend{ELM, GP-RBF, LSTM, SVR, FFNN, GP-DP, True};
\end{axis}
\end{tikzpicture}
\end{adjustbox}
\caption{Daily glucose predictions as a function of time for a child during a specific day.}
\label{fig:graph}
\end{figure*}
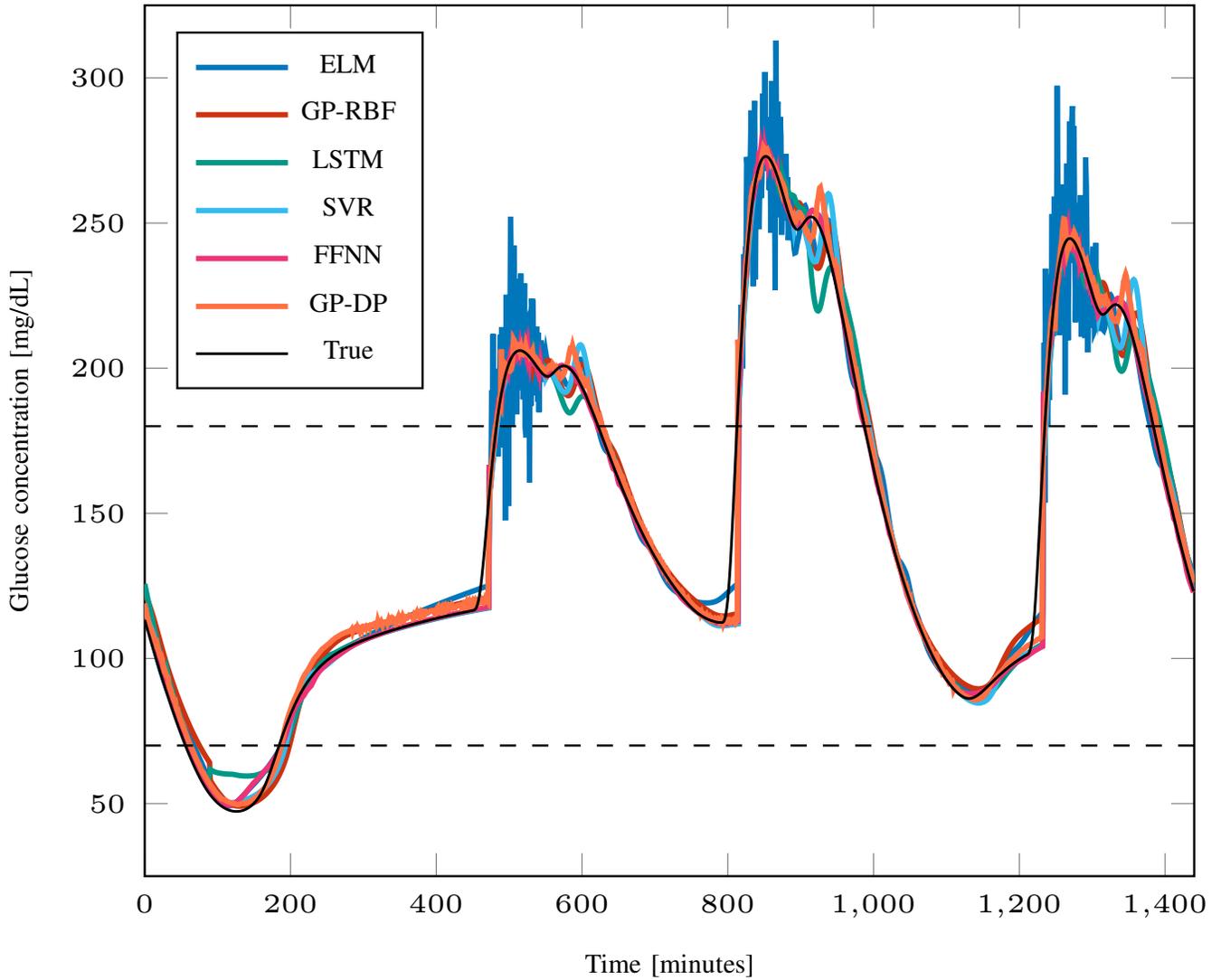

Figure \ref{fig:graph} gives the reader the opportunity to visualize the predictions of the models against the true glucose values for one of the children during a specific day, starting at 0h00 and ending at 23h59. The three peaks in the graph that extend into the hyperglycemia range represent the postprandial rise of glucose.

\begin{table*}
\caption{Performances of glucose prediction models with mean and standard deviation across the children.}
\label{table:results}

\begin{tabularx}{\textwidth}{@{}l||C|*{3}{C}|*{3}{C}|*{3}{C}@{}}
\toprule

\multirow{3}{*}{\textbf{Models}} &  \multirow{3}{*}{\textbf{RMSE}} &    \multicolumn{9}{c}{\textbf{CG-EGA}} \\

& & \multicolumn{3}{c|}{\textit{Hypoglycemia}} & \multicolumn{3}{c|}{\textit{Euglycemia}} & \multicolumn{3}{c}{\textit{Hyperglycemia}} \\ 

& & AP & BE & EP & AP & BE & EP & AP & BE & EP \\

\midrule

\textbf{ELM} &{\textbf{11.24}~(2.74)} &{\textbf{96.01}~(4.24)} &{\textbf{0.13}~(0.11)} &{\textbf{3.85}~(4.30)} &{\textbf{92.95}~(2.94)} &{\textbf{6.04}~(2.90)} &{\textbf{1.01}~(0.58)} &{\textbf{75.95}~(5.09)} &{\textbf{18.58}~(4.07)} &{\textbf{5.47}~(1.84)}\\

\textbf{GP-RBF} &{\textbf{7.84}~(2.51)} &{\textbf{94.82}~(12.23)} &{\textbf{0.07}~(0.08)} &{\textbf{5.11}~(12.26)} &{\textbf{97.15}~(1.43)} &{\textbf{2.74}~(1.42)} &{\textbf{0.12}~(0.05)} &{\textbf{97.53}~(1.62)} &{\textbf{1.95}~(1.29)} &{\textbf{0.51}~(0.42)}\\

\textbf{LSTM} &{\textbf{7.08}~(1.53)} &{\textbf{99.01}~(0.72)} &{\textbf{0.12}~(0.12)} &{\textbf{0.87}~(0.71)} &{\textbf{97.46}~(1.28)} &{\textbf{2.45}~(1.27)} &{\textbf{0.09}~(0.02)} &{\textbf{97.79}~(1.07)} &{\textbf{1.79}~(0.94)} &{\textbf{0.42}~(0.20)}\\

\textbf{SVR} & {\textbf{5.92}~(2.19)}  & {\textbf{99.22}~(0.59)} & {\textbf{0.04}~(0.08)}   & {\textbf{0.74}~(0.55)} & {\textbf{97.29}~(1.55)} & {\textbf{2.61}~(1.54)}   & {\textbf{0.09}~(0.03)}   & {\textbf{97.99}~(1.65)} & {\textbf{1.72}~(1.49)} & {\textbf{0.29}~(0.18)}\\

\textbf{FFNN} &{\textbf{5.43}~(1.84)} &{\textbf{99.09}~(0.51)} &{\textbf{0.09}~(0.07)} &{\textbf{0.81}~(0.49)} &{\textbf{95.29}~(2.15)} &{\textbf{4.30}~(2.09)} &{\textbf{0.40}~(0.17)} &{\textbf{88.60}~(2.25)} &{\textbf{9.32}~(1.87)} &{\textbf{2.08}~(0.54)}\\

\textbf{GP-DP} & {\textbf{5.16}~(1.96)} & {\textbf{99.37}~(1.66)} & {\textbf{0.03}~(0.05)}   & {\textbf{0.61}~(0.45)}    & {\textbf{96.87}~(1.66)} & {\textbf{2.86}~(1.57)} & {\textbf{0.27}~(0.14)} & {\textbf{95.63}~(3.40)} & {\textbf{3.34}~(2.77)} & {\textbf{1.03}~(0.75)}\\

\bottomrule
\end{tabularx}

\begin{flushright}
AP: Accurate Prediction (in \%); BE: Benign Error (in \%); EP: Erroneous Prediction (in \%)
\end{flushright}
\end{table*}

\section{Discussion}


By showing high EP in at least one CG-EGA range, the ELM, the GP-RBF and the FFNN to some extent, do not make safe predictions for the diabetic patients. However, there is no clear winner among the remaining three models. While GP-DP presents the best RMSE and best CG-EGA in the hypoglycemia range, it is surpassed by the LSTM and SVR models in the remaining areas of the CG-EGA, with both being, respectively, the best model in the euglycemia and the best model in the hyperglycemia range. We also note that the GP-DP model has generally higher AP standard deviation values compared to the other models in the CG-EGA. With higher standard deviations, the GP-DP results are shown to be less stable across the diabetic children population, which is not preferable considering potential future use of such predictive models into the whole real diabetic population.

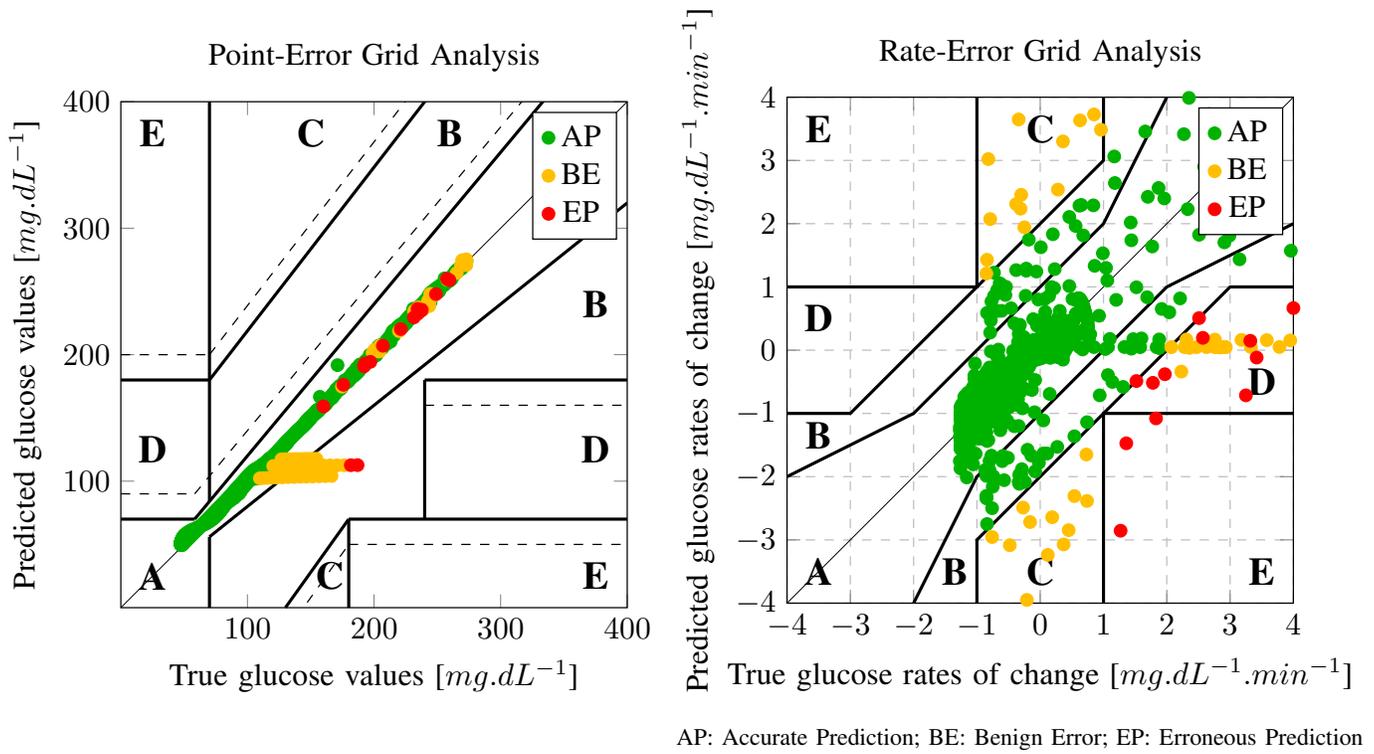
\begin{figure*}
\begin{tikzpicture}
\begin{axis}[
    title={Point-Error Grid Analysis},
    xlabel={True glucose values [$mg.dL^{-1}$]},
    ylabel={Predicted glucose values [$mg.dL^{-1}$]},
    unit vector ratio=1 1 1,
    xmin=0, xmax=400,
    ymin=0, ymax=400,
    ytick={100,200,300,400},
    xtick={100,200,300,400},
]

\addplot[
    color=green!70!black,
    only marks,
    ]
    table {data/CG-EGA/ffnn_child-002_P-EGA_AP.dat};
    
\addplot[
    color=orange!50!yellow,
    only marks,
    ]
    table {data/CG-EGA/ffnn_child-002_P-EGA_BE.dat};
    
\addplot[
    color=red,
    only marks,
    ]
    table {data/CG-EGA/ffnn_child-002_P-EGA_EP.dat};

\addplot[mark=none, black, samples=2, domain=0:400, line width=0.1pt] {x};

\addplot[mark=none, black, samples=2, domain=58.33:400, line width=1pt] {x*6/5};
\addplot[mark=none, black, samples=2, domain=0:58.33, line width=1pt] {70};
\addplot[mark=none, black, samples=2, domain=70:400, line width=1pt, line width=1pt] {x*4/5};
\addplot[mark=none, black, samples=2, line width=1pt] coordinates{(70,0) (70,56)};
\addplot[mark=none, black, samples=2, line width=1pt] coordinates{(70,84) (70,400)};
\addplot[mark=none, black, samples=2, domain=0:70, line width=1pt] {180};
\addplot[mark=none, black, samples=2, domain=70:400, line width=1pt] {x*22/17+89.412};
\addplot[mark=none, black, samples=2, line width=1pt] coordinates{(180,0) (180,70)};
\addplot[mark=none, black, samples=2, line width=1pt] coordinates{(180,70) (400,70)};
\addplot[mark=none, black, samples=2, line width=1pt] coordinates{(240,70) (240,180)};
\addplot[mark=none, black, samples=2, line width=1pt] coordinates{(240,180) (400,180)};
\addplot[mark=none, black, samples=2, domain=130:180, line width=1pt] {x*7/5-182};

\addplot[mark=none, black, samples=2, domain=130:180, style=dashed,] {x*7/5-202};
\addplot[mark=none, black, samples=2, domain=130:180, style=dashed,] coordinates{(180,50) (400,50)};
\addplot[mark=none, black, samples=2, domain=130:180, style=dashed,] coordinates{(240,160) (400,160)};
\addplot[mark=none, black, samples=2, domain=58.33:400, style=dashed,] {x*6/5 + 20};
\addplot[mark=none, black, samples=2, domain=0:58.33, style=dashed,] {90};
\addplot[mark=none, black, samples=2, domain=0:70, style=dashed,] {200};
\addplot[mark=none, black, samples=2, domain=70:400, style=dashed,] {x*22/17+109.412};

\draw (25,25) node[scale=1.25] {\textbf{A}};
\draw (375,240) node[scale=1.25] {\textbf{B}};
\draw (260,375) node[scale=1.25] {\textbf{B}};
\draw (150,375) node[scale=1.25] {\textbf{C}};
\draw (165,25) node[scale=1.25] {\textbf{C}};
\draw (25,125) node[scale=1.25] {\textbf{D}};
\draw (375,125) node[scale=1.25] {\textbf{D}};
\draw (375,25) node[scale=1.25] {\textbf{E}};
\draw (25,375) node[scale=1.25] {\textbf{E}};

\legend{AP, BE, EP}

\end{axis}
\end{tikzpicture}
\hfill
\begin{tikzpicture}

\begin{axis}[
    title={Rate-Error Grid Analysis},
    xlabel={True glucose rates of change [$mg.dL^{-1}.min^{-1}$]},
    ylabel={Predicted glucose rates of change [$mg.dL^{-1}.min^{-1}$]},
    unit vector ratio=1 1 1,
    xmin=-4, xmax=4,
    ymin=-4, ymax=4,
    ytick={-4,-3,-2,-1,0,1,2,3,4},
    xtick={-4,-3,-2,-1,0,1,2,3,4},
    ymajorgrids=true,
    xmajorgrids=true,
    grid style=dashed,
]

\addplot[
    color=green!70!black,
    only marks,
    ]
    table {data/CG-EGA/ffnn_child-002_R-EGA_AP.dat};
    
\addplot[
    color=orange!50!yellow,
    only marks,
    ]
    table {data/CG-EGA/ffnn_child-002_R-EGA_BE.dat};
    
\addplot[
    color=red,
    only marks,
    ]
    table {data/CG-EGA/ffnn_child-002_R-EGA_EP.dat};
    
\addplot[mark=none, black, samples=2, domain=-4:4, line width=0.1pt] {x};

\addplot[mark=none, black, samples=2, line width=1pt] table {
-4 1
-1 1
-1 4
};

\addplot[mark=none, black, samples=2, line width=1pt] table {
-4 -1
-3 -1
1 3
1 4
};

\addplot[mark=none, black, samples=2, line width=1pt] table {
-4 -2
-2 -1
1 2
2 4
};

\addplot[mark=none, black, samples=2, line width=1pt] table {
-2 -4
-1 -2 
2 1
4 2
};

\addplot[mark=none, black, samples=2, line width=1pt] table {
-1 -4
-1 -3
3 1
4 1
};

\addplot[mark=none, black, samples=2, line width=1pt] table {
1 -4
1 -1
4 -1 
};

\draw (-3.5,-3.5) node[scale=1.25] {\textbf{A}};
\draw (-3.5,-1.35) node[scale=1.25] {\textbf{B}};
\draw (-1.35,-3.5) node[scale=1.25] {\textbf{B}};
\draw (0,-3.5) node[scale=1.25] {\textbf{C}};
\draw (0,3.5) node[scale=1.25] {\textbf{C}};
\draw (-3.5,0.5) node[scale=1.25] {\textbf{D}};
\draw (3.5,-0.5) node[scale=1.25] {\textbf{D}};
\draw (-3.5,3.5) node[scale=1.25] {\textbf{E}};
\draw (3.5,-3.5) node[scale=1.25] {\textbf{E}};

\legend{AP, BE, EP}

\end{axis}

\end{tikzpicture}
\begin{flushright}\small{
AP: Accurate Prediction; BE: Benign Error; EP: Erroneous Prediction}
\end{flushright}
\caption{P-EGA and R-EGA of the FFNN model of a child for a specific day. For every prediction, AP, BE and EP markers are computed by combining the P-EGA and the R-EGA into the CG-EGA.}
\label{fig:FFFN-CG-EGA}
\end{figure*}

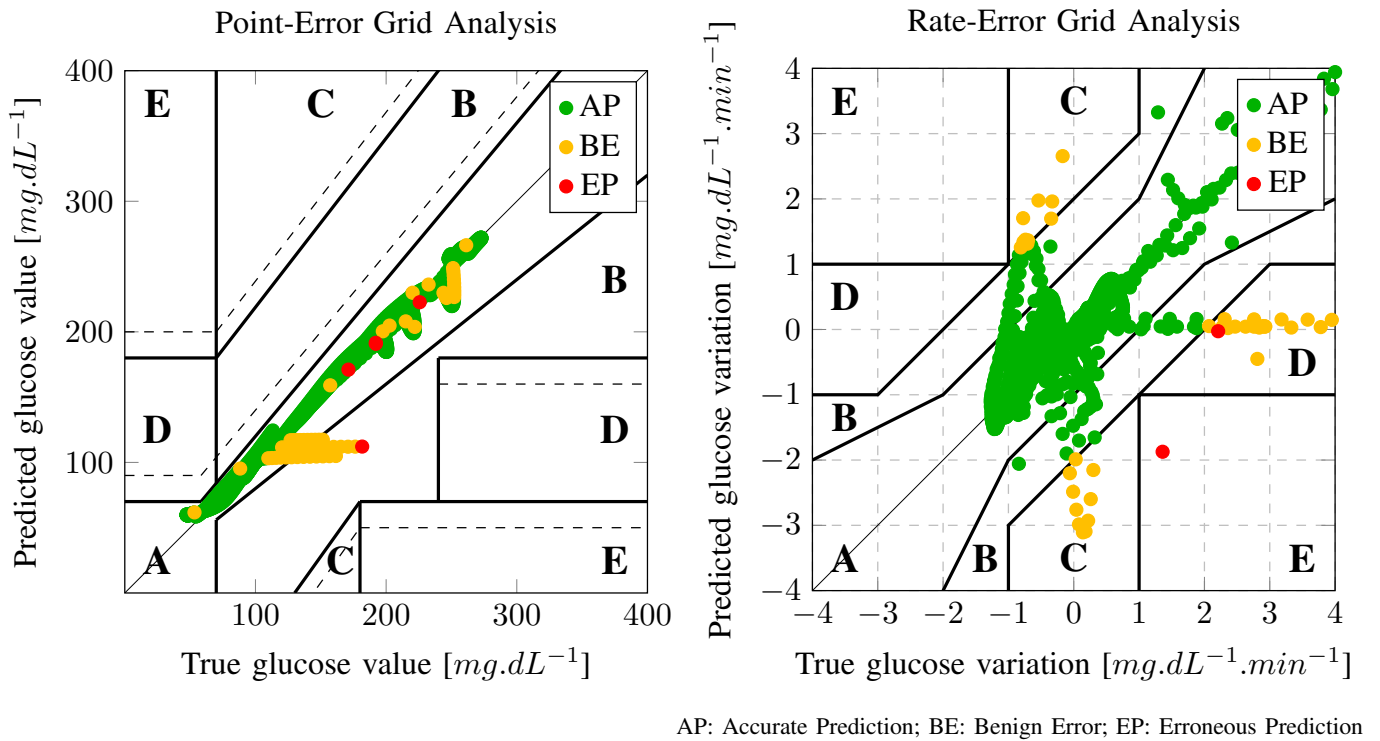
\begin{figure*}
\begin{tikzpicture}
\begin{axis}[
    title={Point-Error Grid Analysis},
    xlabel={True glucose value [$mg.dL^{-1}$]},
    ylabel={Predicted glucose value [$mg.dL^{-1}$]},
    unit vector ratio=1 1 1,
    xmin=0, xmax=400,
    ymin=0, ymax=400,
    ytick={100,200,300,400},
    xtick={100,200,300,400},
]

\addplot[
    color=green!70!black,
    only marks,
    ]
    table {data/CG-EGA/lstm_child-002_P-EGA_AP.dat};
    
\addplot[
    color=orange!50!yellow,
    only marks,
    ]
    table {data/CG-EGA/lstm_child-002_P-EGA_BE.dat};
    
\addplot[
    color=red,
    only marks,
    ]
    table {data/CG-EGA/lstm_child-002_P-EGA_EP.dat};

\addplot[mark=none, black, samples=2, domain=0:400, line width=0.1pt] {x};

\addplot[mark=none, black, samples=2, domain=58.33:400, line width=1pt] {x*6/5};
\addplot[mark=none, black, samples=2, domain=0:58.33, line width=1pt] {70};
\addplot[mark=none, black, samples=2, domain=70:400, line width=1pt, line width=1pt] {x*4/5};
\addplot[mark=none, black, samples=2, line width=1pt] coordinates{(70,0) (70,56)};
\addplot[mark=none, black, samples=2, line width=1pt] coordinates{(70,84) (70,400)};
\addplot[mark=none, black, samples=2, domain=0:70, line width=1pt] {180};
\addplot[mark=none, black, samples=2, domain=70:400, line width=1pt] {x*22/17+89.412};
\addplot[mark=none, black, samples=2, line width=1pt] coordinates{(180,0) (180,70)};
\addplot[mark=none, black, samples=2, line width=1pt] coordinates{(180,70) (400,70)};
\addplot[mark=none, black, samples=2, line width=1pt] coordinates{(240,70) (240,180)};
\addplot[mark=none, black, samples=2, line width=1pt] coordinates{(240,180) (400,180)};
\addplot[mark=none, black, samples=2, domain=130:180, line width=1pt] {x*7/5-182};

\addplot[mark=none, black, samples=2, domain=130:180, style=dashed,] {x*7/5-202};
\addplot[mark=none, black, samples=2, domain=130:180, style=dashed,] coordinates{(180,50) (400,50)};
\addplot[mark=none, black, samples=2, domain=130:180, style=dashed,] coordinates{(240,160) (400,160)};
\addplot[mark=none, black, samples=2, domain=58.33:400, style=dashed,] {x*6/5 + 20};
\addplot[mark=none, black, samples=2, domain=0:58.33, style=dashed,] {90};
\addplot[mark=none, black, samples=2, domain=0:70, style=dashed,] {200};
\addplot[mark=none, black, samples=2, domain=70:400, style=dashed,] {x*22/17+109.412};

\draw (25,25) node[scale=1.25] {\textbf{A}};
\draw (375,240) node[scale=1.25] {\textbf{B}};
\draw (260,375) node[scale=1.25] {\textbf{B}};
\draw (150,375) node[scale=1.25] {\textbf{C}};
\draw (165,25) node[scale=1.25] {\textbf{C}};
\draw (25,125) node[scale=1.25] {\textbf{D}};
\draw (375,125) node[scale=1.25] {\textbf{D}};
\draw (375,25) node[scale=1.25] {\textbf{E}};
\draw (25,375) node[scale=1.25] {\textbf{E}};

\legend{AP, BE, EP}

\end{axis}
\end{tikzpicture}
\hfill
\begin{tikzpicture}

\begin{axis}[
    title={Rate-Error Grid Analysis},
    xlabel={True glucose variation [$mg.dL^{-1}.min^{-1}$]},
    ylabel={Predicted glucose variation [$mg.dL^{-1}.min^{-1}$]},
    unit vector ratio=1 1 1,
    xmin=-4, xmax=4,
    ymin=-4, ymax=4,
    ytick={-4,-3,-2,-1,0,1,2,3,4},
    xtick={-4,-3,-2,-1,0,1,2,3,4},
    ymajorgrids=true,
    xmajorgrids=true,
    grid style=dashed,
]

\addplot[
    color=green!70!black,
    only marks,
    ]
    table {data/CG-EGA/lstm_child-002_R-EGA_AP.dat};
    
\addplot[
    color=orange!50!yellow,
    only marks,
    ]
    table {data/CG-EGA/lstm_child-002_R-EGA_BE.dat};
    
\addplot[
    color=red,
    only marks,
    ]
    table {data/CG-EGA/lstm_child-002_R-EGA_EP.dat};
    
\addplot[mark=none, black, samples=2, domain=-4:4, line width=0.1pt] {x};

\addplot[mark=none, black, samples=2, line width=1pt] table {
-4 1
-1 1
-1 4
};

\addplot[mark=none, black, samples=2, line width=1pt] table {
-4 -1
-3 -1
1 3
1 4
};

\addplot[mark=none, black, samples=2, line width=1pt] table {
-4 -2
-2 -1
1 2
2 4
};

\addplot[mark=none, black, samples=2, line width=1pt] table {
-2 -4
-1 -2 
2 1
4 2
};

\addplot[mark=none, black, samples=2, line width=1pt] table {
-1 -4
-1 -3
3 1
4 1
};

\addplot[mark=none, black, samples=2, line width=1pt] table {
1 -4
1 -1
4 -1 
};

\draw (-3.5,-3.5) node[scale=1.25] {\textbf{A}};
\draw (-3.5,-1.35) node[scale=1.25] {\textbf{B}};
\draw (-1.35,-3.5) node[scale=1.25] {\textbf{B}};
\draw (0,-3.5) node[scale=1.25] {\textbf{C}};
\draw (0,3.5) node[scale=1.25] {\textbf{C}};
\draw (-3.5,0.5) node[scale=1.25] {\textbf{D}};
\draw (3.5,-0.5) node[scale=1.25] {\textbf{D}};
\draw (-3.5,3.5) node[scale=1.25] {\textbf{E}};
\draw (3.5,-3.5) node[scale=1.25] {\textbf{E}};

\legend{AP, BE, EP}

\end{axis}

\end{tikzpicture}
\begin{flushright}\small{
AP: Accurate Prediction; BE: Benign Error; EP: Erroneous Prediction}
\end{flushright}
\caption{P-EGA and R-EGA of the LSTM model of a child for a specific day. For every prediction, AP, BE and EP markers are computed by combining the P-EGA and the R-EGA into the CG-EGA.}
\label{fig:LSTM-CG-EGA}
\end{figure*}

In a completely different perspective, our study shows the usefulness of using the CG-EGA to assess the performances of the models and not relying solely on the RMSE metric. To illustrate this idea, we can compare the results of the FFNN and the LSTM models. As we can see in Table \ref{table:results}, while the FFNN model has the second lowest RMSE, its CG-EGA results in the hyperglycemia range make it clinically unacceptable. If we look into the details of the FFNN's CG-EGA (Figure \ref{fig:FFFN-CG-EGA}), compared to those of the LSTM model (Figure \ref{fig:LSTM-CG-EGA}) from the same child during the same day, we can understand why. In both the P-EGA and the R-EGA figures, a perfect prediction is represented by the line in the middle of the A zones. In both P-EGA grids, while most predictions are in the A zone, we can see that FFNN predictions are closer to true values compared to those made by the LSTM model (this difference is reflected by the difference in RMSE between both models). In the other hand, the R-EGA figures show us that the predicted rates of change of the FFNN model are much more spread out inside the grid. For a model to have overall good CG-EGA results, it needs to perform well in the P-EGA and in the R-EGA grids at the same time. The FFNN model, while being one of the best point predictive models (RMSE or P-EGA), has trouble estimating precisely and consistently the glucose variations. We think that the LSTM model manages to have good CG-EGA results despite not being one of the best RMSE thanks to the inherent nature of the algorithm. RNN, especially those based on LSTM units, make use of the sequential nature of the data to remember important observations to compute coherent predictions.

Finally, the results highlight some limitations of the CG-EGA. First, as it is not usually trained on (given its complex nature), algorithms which are only trained to compute good point predictions (e.g. FFNN)  may not succeed the clinical acceptability test because it involves rates of change predictions. Second, the CG-EGA fails at discriminating models that have more or less the same results (namely LSTM, SVR and GP-DP). We should note that, rather than from the CG-EGA itself, it comes from the common simplification made from it (the AP, BE and EP categories).

\section{Conclusion}

In this paper we studied six of the most trending and promising glucose predictive models. We compared a FFNN, a LSTM RNN, a ELM neural network, a SVR model, and two GP models, one with a RBF kernel and the other with a DP kernel. While the RMSE has been used to measure the accuracy of the predictions, the CG-EGA has been used to provide a measure of the clinical acceptability of the models. The GP-DP model is a novel improvement of GP models, traditionally used with a RBF kernel in the context of glucose prediction. 

The analysis of the results showed that only the LSTM, SVR and GP-DP have overall satisfactory results, each of them having its own strength. In particular, while the GP-DP model presents the best RMSE as well as the best clinical acceptability in the hypoglycemia range, the LSTM and SVR models excel, respectively, in the euglycemia range and in the hyperglycemia range.

Besides, we highlighted the limitations of the evaluation methodology currently used in the field of glucose prediction. While the CG-EGA covers the RMSE weakness by providing a way of evaluating the clinical acceptability of the models, it is not perfect as it cannot be trained on and, given the common simplification used to report the results, cannot help discriminating between the best models.

This study makes us identify new approaches to tackle the problem of predicting future glucose values of diabetic patients such as improving the way we evaluate the models or combining them into a single predictor (through fusion algorithms for instance) which would take advantage of their different strengths.

Finally, in the future, we aim at conducting this study on the other diabetic populations (i.e., adolescents, adults) and for longer prediction horizons (e.g., 60 or 120 minutes) to see if we can generalize our results and findings.

\section*{Acknowledgment}
This  work  is  supported  by  the  "IDI  2017"  project  funded by the IDEX Paris-Saclay, ANR-11-IDEX-0003-02.


\bibliographystyle{IEEEtran}
\bibliography{bibtex.bib}

\end{document}